%% file: bukanko.tex
\documentclass{aa}
\usepackage{txfonts}
\usepackage{natbib}
\usepackage{graphicx}

\bibpunct[, ]{(}{)}{;}{a}{}{,}

\def\hd{$56$~Ari}

\input{common-def}

\begin{document}

\title{The Lorentz force in atmospheres of CP stars: 56~Arietis}

\author{
D. Shulyak\inst{1} \and
O. Kochukhov\inst{2} \and
G. Valyavin\inst{3} \and
B.-C. Lee\inst{4} \and 
G. Galazutdinov\inst{5} \and
K.-M. Kim\inst{4} \and 
Inwoo Han\inst{4} \and 
T.~Burlakova\inst{6}
}

\offprints{D. Shulyak, \\
\email{denis.shulyak@gmail.com}}

\institute{
Institut f\"ur Astronomie, Universit\"at Wien, T\"urkenschanzstra{\ss}e 17, 1180 Wien, Austria \and
Department of Physics and Astronomy, Uppsala University, Box 515, 751 20, Uppsala, Sweden \and
Observatorio Astron\'{o}mico Nacional SPM, Instituto de Astronom\'{i}a, Universidad Nacional Aut\'{o}noma de M\'{e}xico, Ensenada, BC, M\'{e}xico \and
Korea Astronomy and Space Science Institute, 61-1, Whaam-Dong, Youseong-Gu, Taejeon, 305-348, Rep. of Korea \and
Department of Physics and Astronomy, Seoul National University, Gwanak-gu, Seoul 151-747, Rep. of Korea \and
Special Astrophysical Observatory, Russian Academy of Sciences, Nizhnii Arkhyz, Karachai Cherkess Republic, 369167, Russia
}

\date{Received / Accepted}

\abstract
{The presence of electric currents in the atmospheres of magnetic chemically peculiar (mCP) stars
could bring important theoretical constrains about the nature and evolution of magnetic field in these stars.
The Lorentz force, which results from the interaction between the magnetic field and
the induced currents, modifies the atmospheric structure and induces characteristic rotational 
variability of pressure-sensitive spectroscopic features, that can be analysed using phase-resolved
spectroscopic observations.}
{In this work we continue the presentation of results of the magnetic pressure studies in mCP stars
focusing on the high-resolution spectroscopic observations of Bp star \object{\hd}. 
We have detected a significant variability of the \halpha, \hbeta, and \hgamma\ spectral lines 
during full rotation cycle of the star. Then these observations are interpreted in the framework of the model atmosphere
analysis, which accounts for the Lorentz force effects.}
{We used the \llm\ stellar model atmosphere code for the calculation of the magnetic pressure effects
in the atmosphere of \hd\, taking into account realistic chemistry of the star and accurate computations
of the microscopic plasma properties. The \synth\ code was employed to simulate phase-resolved variability 
of Balmer lines.}
{We demonstrate that the model with the outward-directed Lorentz force
in the dipole$+$quadrupole configuration is likely to reproduce the observed 
hydrogen lines variation.
These results present strong evidences for the presence of non-zero
global electric currents in the atmosphere of this early-type magnetic star.}
{}
\keywords{stars: chemically peculiar -- stars: magnetic fields -- stars: atmospheres -- stars: individual: 56~Ari}

\maketitle

\section{Introduction}
The atmospheres of magnetic chemically peculiar (mCP) stars display the
presence of global magnetic
fields ranging in strength from a few hundred G up to several tens of
kG \citep{landstreet2001}, with the global configurations well represented by 
dipolar or low-order multipolar components \citep{bagnulo} and
that are likely stable during significant time intervals. 
The stability of the atmospheres against strong
convective motions and the existence of large scale magnetic fields provide a unique conditions
for the study of secular evolution of global cosmic magnetic fields and other
dynamical processes which may take place in the magnetized plasma.
In particular, the slow variation of the field geometry and strength
changes the pressure-force balance in the atmosphere via the
induced Lorentz force, that makes it possible to detect it observationally
and establish a number of important constraints on the plausible scenarios 
of the magnetic field evolution in early-type stars.

Among the known characteristics of mCP stars, the variability
of hydrogen Balmer lines is poorly understood. For some of stars it can be connected with
inhomogeneous surface distribution of chemical elements, possible temperature variations
and/or stellar rotation \citep[see, for example, discussion in][]{lehmann-2007}.
At the same time, some of the magnetic stars demonstrate the characteristic shape of
the Balmer line variability that can not be simply described by temperature
or abundance effects. For example, \citet{kroll} showed that at least part of 
the variability detected in several mCP stars can be attributed to the pressure effects, indicating the presence 
of a non-zero Lorentz force in their atmospheres.

Different atmosphere models with the Lorenz force were considered by
several authors \citep[see review in][]{valyavin}. In this study
we follow approaches given by \citet{valyavin} considering the
problem in terms of induced atmospheric electric currents interacting
with magnetic fields. The authors predicted that the amplitude of the variations
in hydrogen Balmer lines seen in real stars can be described if one assumes strong electric currents
flowing in upper atmospheric layers of these objects. 
Later on, \citet{lorentz-2007} (\paperone\, hereafter) extended and improved
this model to more complicated field geometries and accurate treatment of magnetized 
plasma properties. Their first direct implementation of new model atmospheres to the 
analysis of the hydrogen spectra of one of the brightest mCP star $\theta$~Aur
showed that their rotational modulation can be induced by the Lorentz force effect, which is not directly connected to the temperature or abundance variation across the stellar surface.
The precise analysis of the longitudinal magnetic field variation and subsequent
modeling of the magnetic pressure for every observed rotational phase of the star
allowed us to constrain the magnitude and the direction of the Lorentz force.
In particular, the outward-directed Lorentz force (i.e. directed outside the stellar interior
along radius) was found to provide the best fit to observations, in combination
with rather strong induced \emf\ (electro-magnetic force) of about $1\times10^{-11}$~CGS units
which was found to play an important role in the overall hydrostatic structure of the stellar atmosphere.

The knowledge of the direction and the strength of the Lorentz force is important for the understanding
of physical mechanisms that are responsible for such strong surface currents and their interaction
with global, large-scale magnetic fields (see discussion in \paperone\, for more details).
Taking this into account and following the pioneering work by \citet{kroll}, we have
initiated a new spectroscopic search of hydrogen line variability in a number of magnetic
main-sequence stars \citep{valyavin-2005}.

In this paper we present the phase-resolved high-resolution observations
of one of the weak-field ($|$\bz$|$\,$<$\,500~G) mCP star \hd\ (HD~19832). 
We detected significant variation of the Balmer line profiles and
interpreted it in terms of the non-force-free magnetic field configuration.

The overview of observations will be presented in the next section. Then,
in Sect.~3 we will give a short description of the model used to simulate 
the effects of the magnetic pressure. Main results will be summarized in
Sect.~4 with conclusions and discussion given in Sect.~5 and~6 respectively.

\section{Observations}
Observations of \hd\ were carried out with the BOES echelle spectrograph
installed  at the $1.8$\,m telescope of the Korean Astronomy and Space Science
Institute. The spectrograph and observational procedures are described by
\citet{kim} and by \citet{kim2}. The instrument is a moderate-beam, fibre-fed
high-resolution spectrometer
which incorporates 3 STU Polymicro fibres of $300$, $200$, and $80$~$\mu m$ core
diameter (corresponding spectral resolutions are
$\lambda/\Delta\lambda = 30\,000$, $45\,000$, and $90\,000$ respectively).
The medium resolution mode was employed in the present study. Working
wavelength range is from $3500$\AA\, to $10\,000$\AA.

Seventeen spectra of the star were recorded in the course of $10$ observing
nights from 2004 to 2006. Typical exposure times of a few
minutes allowed to achieve $S/N\approx 250$--300.
Table~\ref{tab:obs} gives an overview of our observations.
Throughout this study we implement ephemeris derived by \citet{adelman-ephem}
using linear changing period model:
\begin{equation}
JD = 2434322.354 + \frac{0.7278883}{1 + 0.7278883 \cdot S \cdot (t-t_{\rm 0})},
\end{equation}
where $t_{\rm 0} = 2434322.354$ and $S = -1.35 \cdot 10^{-9}$.

Details related to spectral data reduction and processing, also
study of the spectrograph's stability are presented in
\citet{lorentz-2007}, and we don't describe it here. Accuracy of the
continuum normalization around Balmer lines is estimated 
to be approximately $0.2-0.3$\%.

\input{tables/tab-obs}

\section{Model}
\subsection{General equations and approximations}
In this section we follow the approach and methods outlined previously in \paperone.
However, for the sake of explanation, we find it useful to state here some of the 
general assumptions used in the modeling procedure:

%
\begin{enumerate}
\item
The stellar surface magnetic field is axisymmetric and is dominated by
dipolar or dipole+quadrupolar component in all atmospheric layers.
\item
The induced \emf\, has only an azimuthal component, similar to that
described by \citet{wrubel}, who considered decay of the global stellar
magnetic field. In this case the distribution of the surface
electric currents can be expressed by the Legendre polynomials
$P^1_n(\mu)$, where $n = 1$ for dipole,
$n = 2$ for quadrupole, etc., and $\mu=\cos\theta$ is the cosine of the co-latitude
angle $\theta$ which is counted in the coordinate system connected
to the symmetry axis of the magnetic field.
\item
The atmospheric layers are assumed to be in static equilibrium and no
horizontal motions are present.
\item
Stellar rotation, Hall's currents, ambipolar diffusion and other dynamical
processes are neglected.
\end{enumerate}

Taking these approximations into account and using Maxwell equation for field vectors and Ohm's law, 
one can write the hydrostatic equation in the form
\begin{equation}
\frac{\partial P_{\rm total}}{\partial r} =
-\rho g \pm \frac{1}{c} \lambda_{\rm \perp}
\sum_n c_n P^1_n(\mu) \sum_n B^{(n)}_{\rm \theta}=-\rho g_{\rm eff}.
\label{eq:hydro}
\end{equation}
Obtaining this equation we used the superposition principle for field
vectors and the solution of Maxwell equations for each of the multipolar
components following \citet{wrubel}. We also suppose that
$\vec{E}\perp\vec{B}$. 
Here $c_n$ represents the effective electric field generated by
the $n$-th magnetic field component at the stellar magnetic equator
and $B_{\rm \theta}$ is the horizontal field component.
The signs ``+'' and ``--'' refer
to the outward- and inward-directed Lorentz forces, respectively.

We note that the values of $c_n$ are free parameters to be found
by using our model. These values represent the fundamental characteristics that
can be used for building self-consistent models of the
global stellar magnetic field geometry and its evolution. Thus, an indirect
measurement of these parameters via the study of the Lorentz force is of
fundamental importance for understanding the stellar magnetism.

Calculation of the electric conductivity $\lambda_{\rm \perp}$ is
carried out using the Lorentz collision model where
only binary collisions between particles are allowed which is a good
approximation for a low density stellar atmosphere plasma.
The detailed description and basic relationships of this approach are given in
\paperone.

From the Eq.~(\ref{eq:hydro}) it is seen that, in the presence of electric currents
and magnetic field, the rotation of a star can produce phase-dependent Lorentz-force term
(due to a variation of $B^{(n)}_{\theta}$ and $\lambda_{\rm \perp}$), which will in turn modify the hydrostatic structure 
of the atmosphere, manifesting itself as a variation of pressure-sensitive spectral lines.

\subsection{Model atmospheres with Lorentz force}
Our calculations were carried out with the stellar model atmosphere
code \llm\, developed by \citet{llm}.
At each iteration the code calculates electric conductivity in all
atmospheric layers using all available charged and neutral plasma particles.
The conductivity is then used to evaluate the magnetic
contribution to the magnetic gravity and to execute temperature and mass correction
procedure.

The input parameters for the calculation of magnetic pressure are the direction of the Lorentz force
(inward- or outward-directed), \emf\, at the stellar equator, mean surface magnetic field modulus \bs,
and the product of the two sums in Eq.~(\ref{eq:hydro}) containing contribution 
from all considered multipolar components for every single rotational phase of the star.

As can be seen from Eq.~(\ref{eq:hydro}), there is some critical value of $c_n$
that may produce unstable solution in the case of the outward-directed Lorentz force.
Such models cannot be considered in the hydrostatic equilibrium approximation
introduced above and were deemed non-physical
in our calculations. Thus, for each set of models, we restricted $c_n$ values
to ensure static equilibrium.

The atomic line list was extracted from the VALD database \citep{vald1,vald2}, 
including all lines originating from the predicted and observed energy levels. 
This line list was used as input
for the lines opacity calculation in the \llm\, code.

\section{Numerical Results}
\subsection{Model atmosphere parameters of \hd}
\begin{figure*}[!t]
\includegraphics[angle=90,width=\hsize]{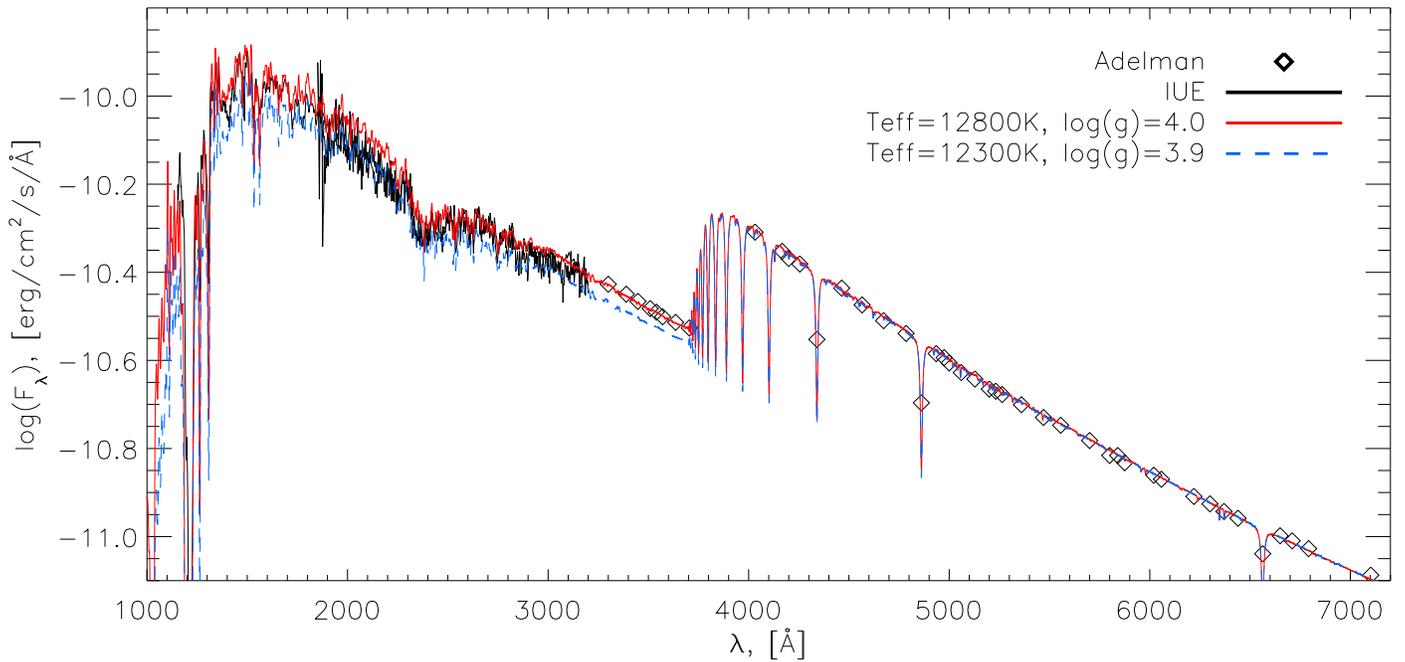}
\caption{Comparison of the observed and computed spectral energy distributions of \hd.
Theoretical models correspond to $\teff=12\,300$~K, $\logg=3.9$ and $\teff=12\,800$~K, $\logg=4.0$. The model fluxes have
been convolved with an $FWHM$\,=\,10\,\AA\, Gaussian kernel for a better view.}
\label{fig:sed}
\end{figure*}

The model atmosphere parameters, $\logg$ and $\teff$, were determined using
theoretical fit of the hydrogen Balmer lines and spectral energy distribution.
It was constructed by combining the average of the optical spectrophotometric
scans obtained by \citet{adelman-phot} and low dispersion UV spectrograms from
the IUE INES\footnote{{\tt http://ines.ts.astro.it/}} database.
For modeling the hydrogen \hbeta\ and \hgamma\ line profiles we used
the mean spectrum of \hd\, averaged other all 17 observed
rotational phases. The projected rotational velocity \vsini\,=\,160~\kms\, 
was taken from \citet{hatzes}. \citet{ryabchik} used the same value in a more recent Doppler imaging study. Individual abundances of several chemical elements, listed in Table~\ref{tab:abn}, were determined as described below (Sect.~\ref{sec:abn}).

\input{tables/tab-abn}

Synthetic Balmer line profiles were calculated using the \synth\, program \citep{synth}, 
which incorporates recent improvements in the treatment of the hydrogen 
line opacity \citep{barklem}.
The stellar energy distribution and Balmer lines are approximated best
with the following parameters: $\teff=12\,800\pm300$~K,
$\logg=4.0\pm0.05$. Note that such high accuracy of the determined
parameters is just an internal accuracy obtained from our technique, which
we used to fit the data. Real parameters
may be slightly different from the obtained ones due to various systematic error sources, 
but this does not play a significant role in our study.

\begin{figure}[h]
\includegraphics[width=\hsize]{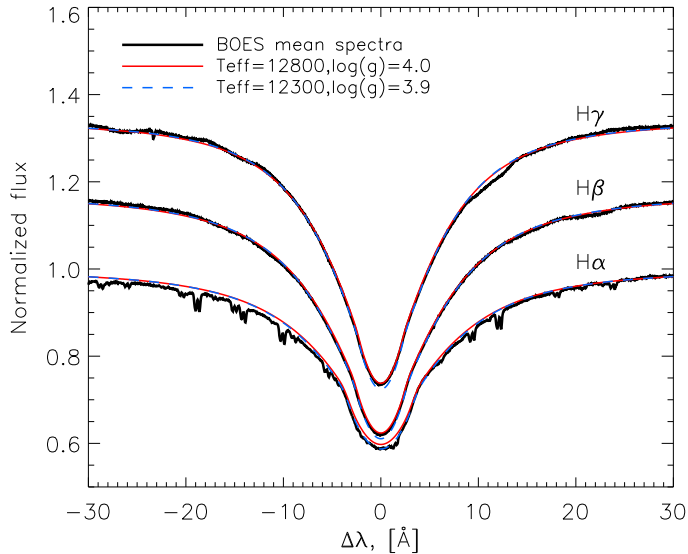}
\caption{Comparison of the observed and computed \halpha, \hbeta, and \hgamma\ line profiles.}
\label{fig:hlines}
\end{figure}

Comparisons of the observations and model predictions are presented
in Figs.~\ref{fig:sed} and~\ref{fig:hlines}. 
We transformed Adelman's spectrophotometric observations to the absolute units
following \citet{lipski-stepien}. Since the absolute calibration of IUE fluxes around their red end
have substantial uncertainties 
\citep[see][]{lipski-stepien},
we scaled the IUE fluxes by $\approx$\,10\% to match the Adelman's data in the near-UV region. This correction is comparable to the offset between alternative flux calibrations suggested for the IUE data \citep{iue}. Applying the correction ensured that the IUE spectra are smooth continuation of the optical
spectrophotometry. Then the theoretical fluxes can be adjusted to fit
the combined set of observations. 
\citet{lipski-stepien} also noted the discrepancy between
observations in visual and UV, but did not do any attempts to correct it. This is why their
final effective temperature of \hd\, was found to be $\teff=12\,250$\,K resulting from the
fact that ignoring the obviously spurious offset between the observed datasets one needs a
lower $\teff$ to fit the IUE fluxes simultaneously with Adelman's fluxes redward 
of the Balmer jump.
As an example we also show in Fig.~\ref{fig:sed} theoretical flux obtained 
from the $\teff=12\,300$\,K, $\logg=3.9$ model which
fits reasonably well the hydrogen Balmer lines (see Fig.~\ref{fig:hlines}), 
but fails to reproduce either Adelman's or IUE data.

Photometric observations in the Str\"omgren and UBV systems also point to the higher $\teff$ and $\logg$ of \hd.
For instance, comparing theoretically computed color-indices with observations of \citet{hauck} and
\citet{nicolet} we find that observed photometric parameters ($b-y=-0.052$, $c_{\rm 1}=0.55$,
$B-V=-0.12$, $U-B=-0.42$) are best fitted with $\teff=12\,800$\,K, $\logg=4.0$ model
($b-y=-0.052$, $c_{\rm 1}=0.60$, $B-V=-0.12$, $U-B=-0.43$) rather than more cooler $\teff=12\,300$\,K, $\logg=3.9$
model ($b-y=-0.048$, $c_{\rm 1}=0.67$, $B-V=-0.12$, $U-B=-0.38$).

Finally, recent studies by \citet{zeeman_paper1} and \citet{zeeman_paper2} showed that the effects of Zeeman
splitting and polarized radiative transfer on the model atmosphere structure and shapes of hydrogen 
line profiles are less than $0.1$\% for magnetic field intensities
around $1$~kG, so they can be safely neglected in the present investigation.

\begin{figure}[h]
\includegraphics[width=\hsize]{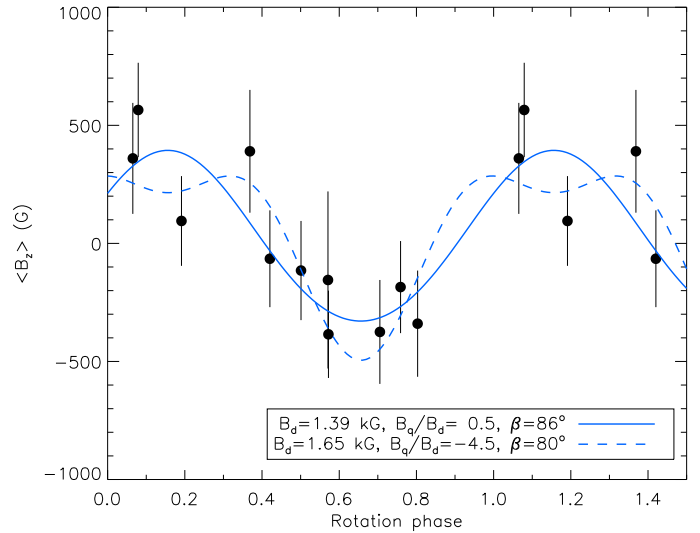}
\caption{Comparison of the longitudinal field observations of \hd\ (symbols)
and the model curves for $i=60\degr$ and two extreme values of the $B_{\rm q}/B_{\rm d}$ parameter (lines).}
\label{fig:bz}
\end{figure}

\subsection{Magnetic-field geometry}
\label{sec:mag-geom}
To calculate the Lorentz force effects, it is essential to specify the
magnetic-field geometry (see Eq.~(\ref{eq:hydro})). For this purpose we
made use of the longitudinal magnetic field measurements obtained by 
\citet{borra} using \hbeta\ photopolarimetric technique. 
The authors observed a smooth single-wave \bz\ variation with rotation phase
and concluded that it is probably caused by a dipole inclined to the
rotation axis of the star.

Generalizing this work, we have approximated the magnetic-field topology of
\hd\ by a combination of the dipole and axisymmetric quadrupole magnetic
components. We also assumed that the symmetry axes of the dipole and
quadrupole magnetic fields are parallel. Thus, the magnetic model parameters
include the polar strength of the dipolar component
$B_{\rm d}$, relative contribution of the quadrupole field
$B_{\rm q}/B_{\rm d}$, magnetic obliquity $\beta$,
and inclination angle $i$ of the stellar rotation axis with respect to
the line of sight. 
The last parameter can be estimated from the usual oblique rotator relation 
connecting stellar radius, rotation period, and \vsini. 
Employing the recently revised Hipparcos parallax of \hd, $\pi=6.49\pm0.76$~mas
\citep{hipp}, $\teff$\,=\,12800~K and bolometric correction $BC=-0.74$
determined by \citet{lipski-stepien}, we found a stellar radius $R=2.8\pm0.4$~$R_\odot$ and inclination angle
$i=57\pm13\degr$. However, the previous Doppler imaging studies of
\citet{hatzes} and \citet{ryabchik} favoured the value of $i=70\degr$. Consequently we decided to explore the model parameters for the entire range
of $i=50$--70\degr.

The remaining free parameters of the magnetic field model
were determined with the least-square fit of the observed \bz\ variation
\citep{borra}. 
Compared to $\theta$~Aur investigated in \paperone, the longitudinal magnetic field curve of \hd\ is poorly defined due to large observational errors and a relatively small number of measurements.
%
%
In this situation 
the acceptable solutions for the magnetic field geometry
fall in a rather wide range: from $B_{\rm q}/B_{\rm d}=0.5$ to $B_{\rm q}/B_{\rm d}=-5$ for
$i=50^\circ$, to $B_{\rm q}/B_{\rm d}=-4.5$ for $i=60^\circ$, and to $B_{\rm q}/B_{\rm d}=-4$ 
for $i=70^\circ$ (Fig.~\ref{fig:bz}). The corresponding dipolar field strength range is $B_{\rm d}=1.3$--1.8~kG and magnetic obliquity is $\beta=70$--90\degr. The lowest $\chi^2_\nu\equiv\chi^2/\nu$ was found between 
$B_{\rm q}/B_{\rm d}=[-1.5,-3]$ with the average value $\chi^2_\nu\approx0.66$ 
while $\chi^2_\nu\approx0.71$ for other geometries: with such a small difference 
we conclude that all the considered magnetic field geometries are 
equally possible.

\subsection{Effects of horizontal abundance distribution}
\label{sec:abn}

We used spectrum synthesis calculations to  access chemical properties of the
atmosphere of \hd. Abundances of He, Al, Mg, Si, and Fe were determined by fitting
\synth\ spectra to the average observations of the star. Results of this analysis,
summarized in Table~\ref{tab:abn}, indicate that He, Al and Mg are deficient in the atmosphere of \hd\
with respect to the solar chemical composition. Fe is moderately overabundant while Si
is strongly enhanced. Abundance of Cr cannot be determined reliably, but since no
prominent \ion{Cr}{ii} lines are present in the spectrum of \hd, we concluded that its
overabundance is most likely smaller than that of Fe.

In addition to the mean abundances analysis we interpreted variation of the phase-resolved spectra 
using the Doppler imaging (DI) technique \citep{abund_di}. Details of this work will be presented in a 
separate publication. In the present paper we are only concerned with the possible effect of the 
horizontal abundance variations on the hydrogen line profiles. Effects of the possible vertical 
stratitification of chemical elemets are ignored because for the purpose of our study they will not differ much 
from the effects of horizontal abundance inhomogeneities. By deriving surface-resolved abundances we effectively 
account for the line opacity variation which would be introduced by chemical stratification.

Among the elements mentioned above only Mg and Si show strong line profile variations. 
Therefore, we reconstructed abundance maps of Si using the red doublet \ion{Si}{ii} 6347, 6371~\AA\ and Mg 
using the 4481~\AA\ line. The modeling of the latter region also included the \ion{He}{i} 4471~\AA\ line, 
which allowed us to estimate potential influence of the He abundance variation on the model atmosphere 
structure and hydrogen line profile behaviour.

\begin{figure}
\includegraphics[height=\hsize,angle=90]{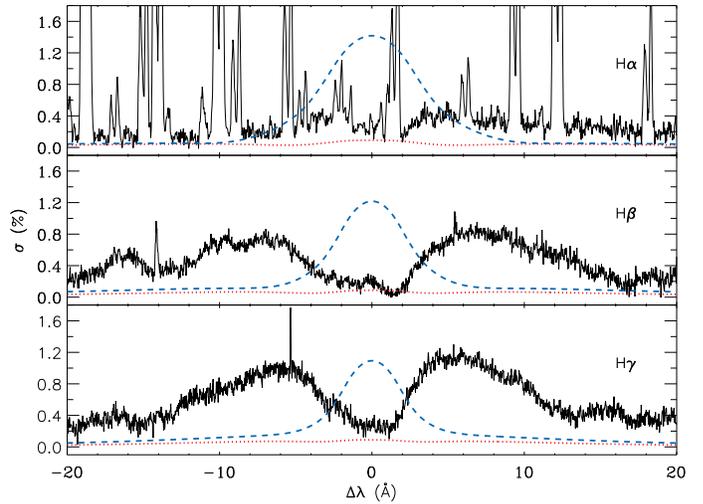}
\caption{The standard deviation $\sigma$ at the H$\alpha$,
H$\beta$, and H$\gamma$ lines. Observations are shown by solid lines. 
Dotted line -- variability due to inhomogeneous
element distributions in \hd. The additional effect of the hypothetical extreme horizontal Fe
gradient is shown by the dashed line (see text for explanations).}
\label{fig:sigma-abn}
\end{figure}

Analysis of Si, He, and Mg demonstrated that horizontal abundance 
inhomogeneities give a negligible contribution to the hydrogen line profile variation.
We have computed model atmospheres with the surface-averaged abundances for each rotational phase of the star. 
Figure~\ref{fig:sigma-abn} shows that the resulting standard deviation of the synthetic profiles is minute compared 
to the variation observed in the line wings of \halpha, \hbeta, and \hgamma.

Much larger abundance gradients of the iron peak elements are required to induce a
noticeable modulation of the hydrogen line profiles during rotation cycle. The dashed
line in Fig.~\ref{fig:sigma-abn} shows results of the numerical experiment where we
have assumed that Fe has the same horizontal distribution as Si but with 10 times
higher contrast. This calculation contradicts the actual observations since the Fe
lines in \hd\ are varying weakly and differently from Si, but it represents a useful
illustration of the impact of chemical inhomogeneities. One can see that the metal
abundance spots lead to the variation of the depth of hydrogen line cores, while in
\hd\ we see changes in the line wings.  Thus, following \paperone\ and \citet{kroll},
we are led to conclude that the chemical spots can not contribute to the observed
Balmer line wing variations.

\subsection{The Lorentz force}
To predict the
phase-resolved variability in hydrogen line profiles both the
inward- and outward-directed Lorentz forces
are examined through the model atmosphere calculations.
The actual magnetic input parameters of
computations with the \llm\, code include the sign of the Lorentz force,
magnetic field modulus $B$, and
the product of two sums $\sum c_n P^1_n \sum B^{(n)}_{\rm\theta}$ (see Eq.~\ref{eq:hydro}).
We take the last two parameters to be disk-averaged at the individual rotation phases
incorporating them in 1D stellar atmosphere code. The corresponding phase curves of the magnetic parameters relevant for our calculations are illustrated in Fig.~\ref{fig:mag}.


Taking into account the large variety of possible solutions for the surface magnetic field 
geometry (see Sect.~\ref{sec:mag-geom}) we calculated a number of model grids 
with the outward- and inward-directed Lorentz force
and $B_{\rm q}/B_{\rm d}$ ranging from $0.5$ to $-4.5$ with a step of~$0.5$.

The calculations showed that, in order to reproduce the amplitudes of the observed standard 
deviations due to the profile variations, the effective electric field should be in the range 
$c_1=1\times10^{-10}-5\times10^{-10}$~CGS in the case of the inward-directed Lorentz force
and $c_1=1.2\times10^{-11}-1\times10^{-10}$~CGS in the case of the outward-directed Lorentz force.
Under the assumption of a purely dipolar configuration these values are: $c_1=5\times10^{-10}$~CGS
for inward-directed and $c_1=7.5\times10^{-11}$~CGS for outward-directed Lorentz forces.


\begin{figure*}
\includegraphics[width=0.33\hsize]{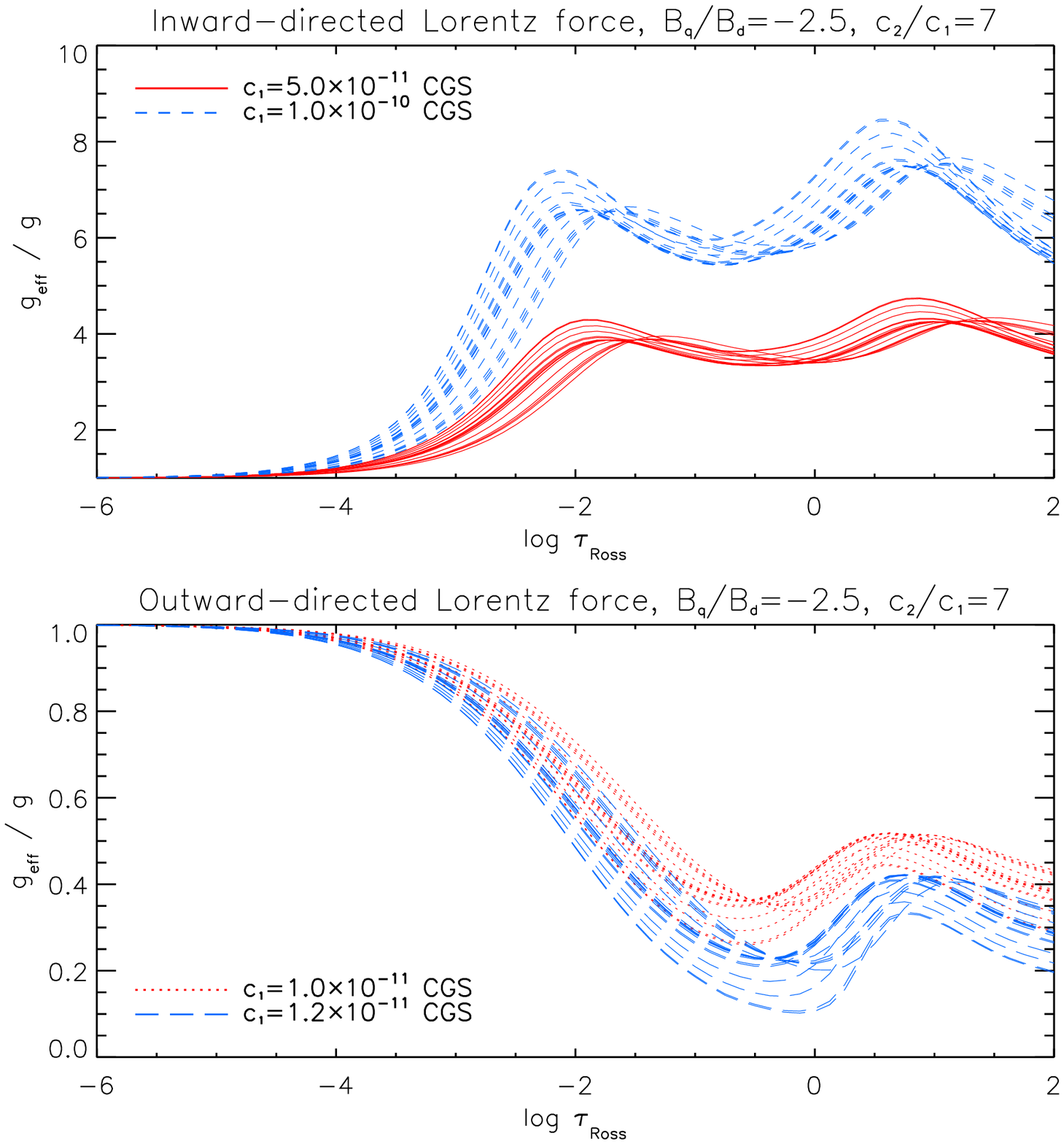}
\includegraphics[width=0.33\hsize]{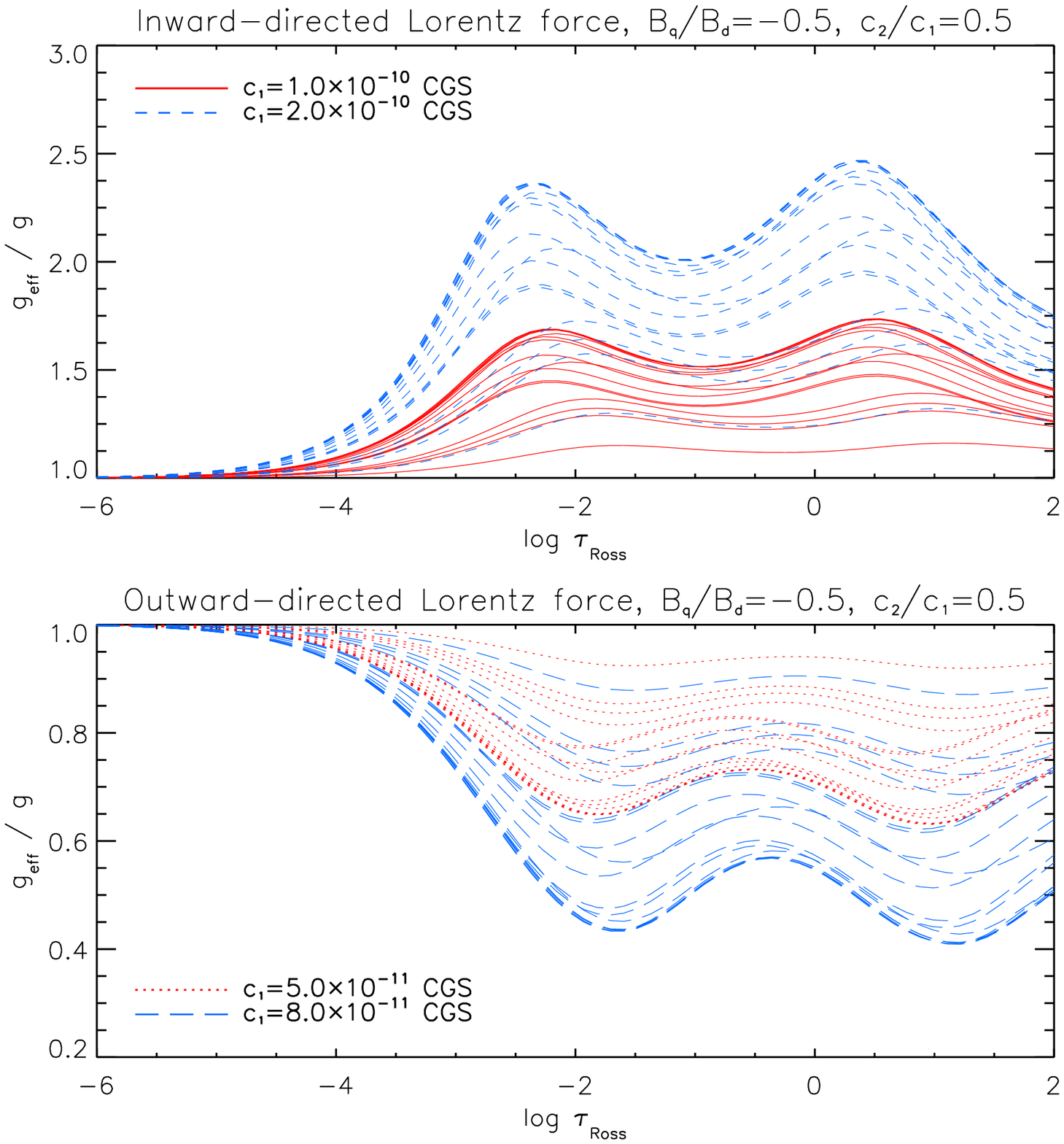}
\includegraphics[width=0.33\hsize]{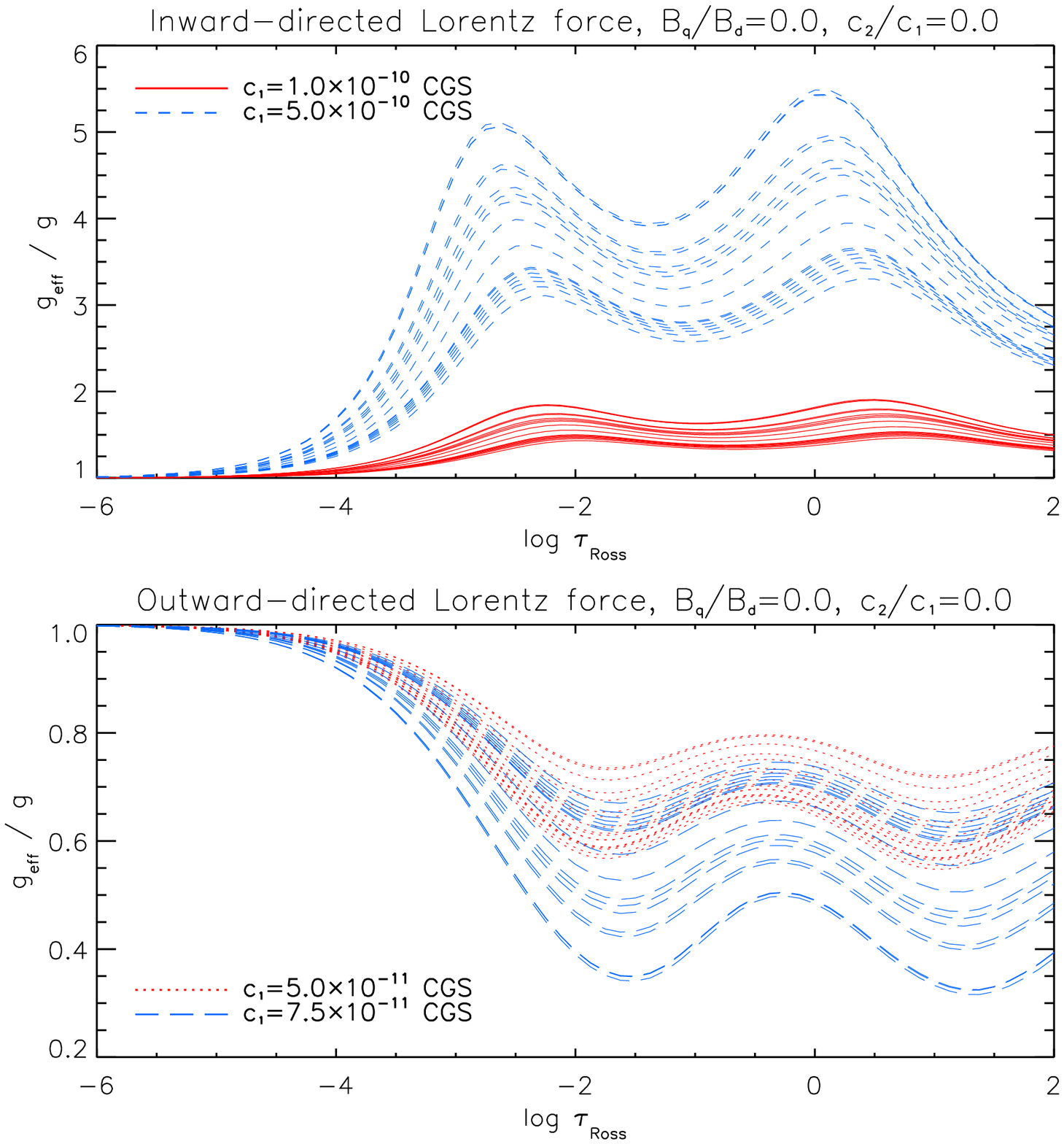}\\
\includegraphics[height=0.33\hsize,angle=90]{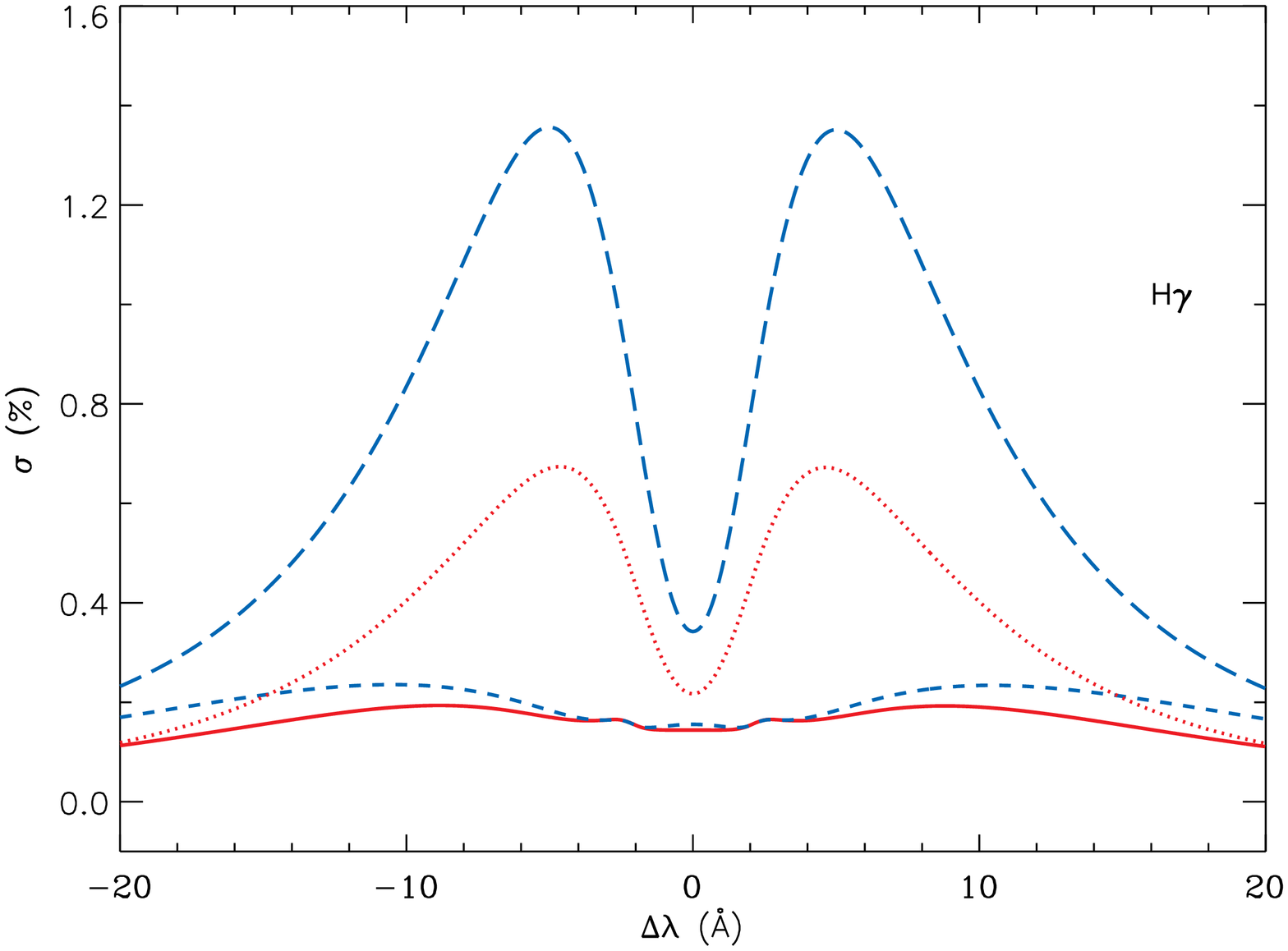}
\includegraphics[height=0.33\hsize,angle=90]{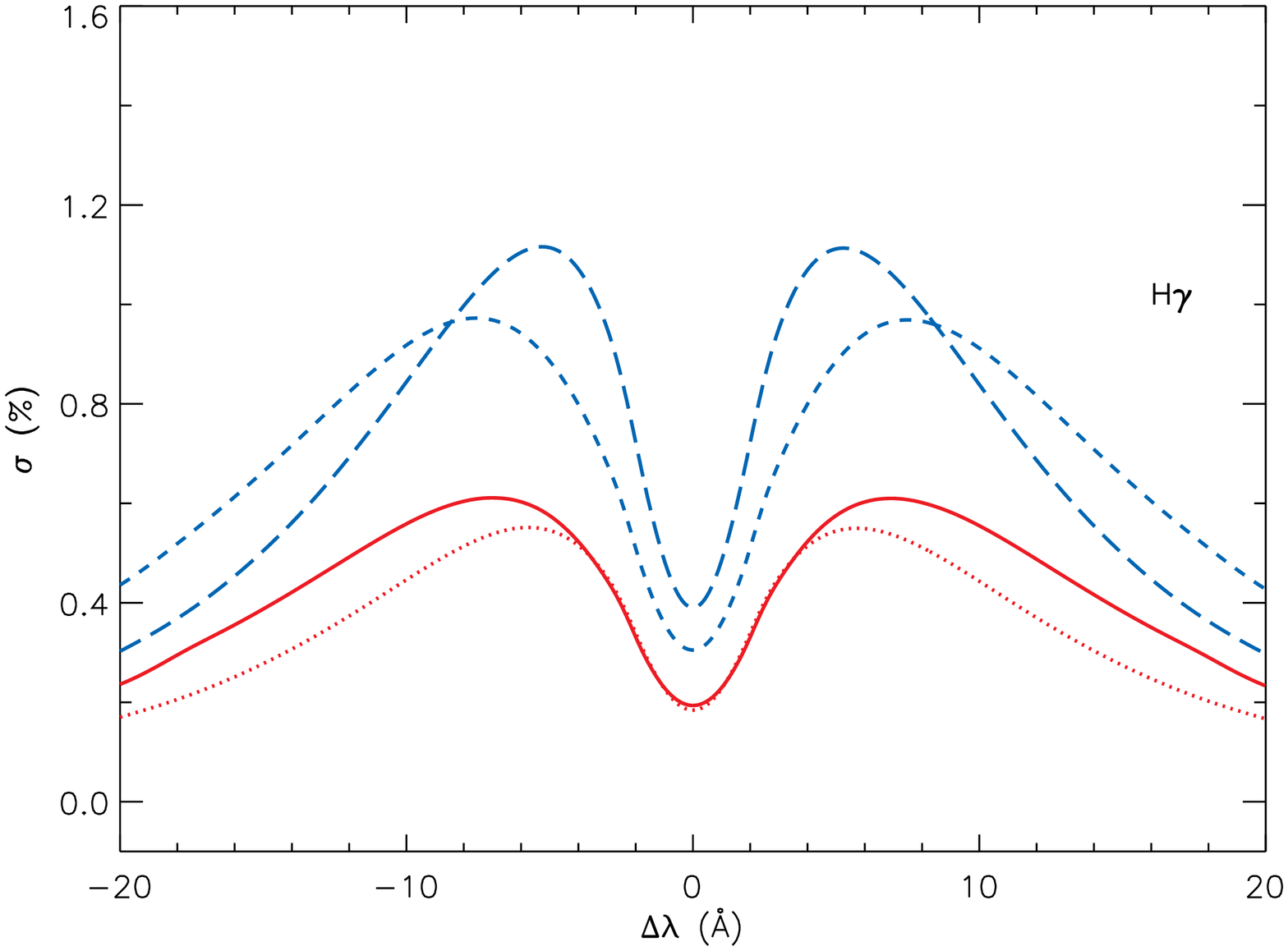}
\includegraphics[height=0.33\hsize,angle=90]{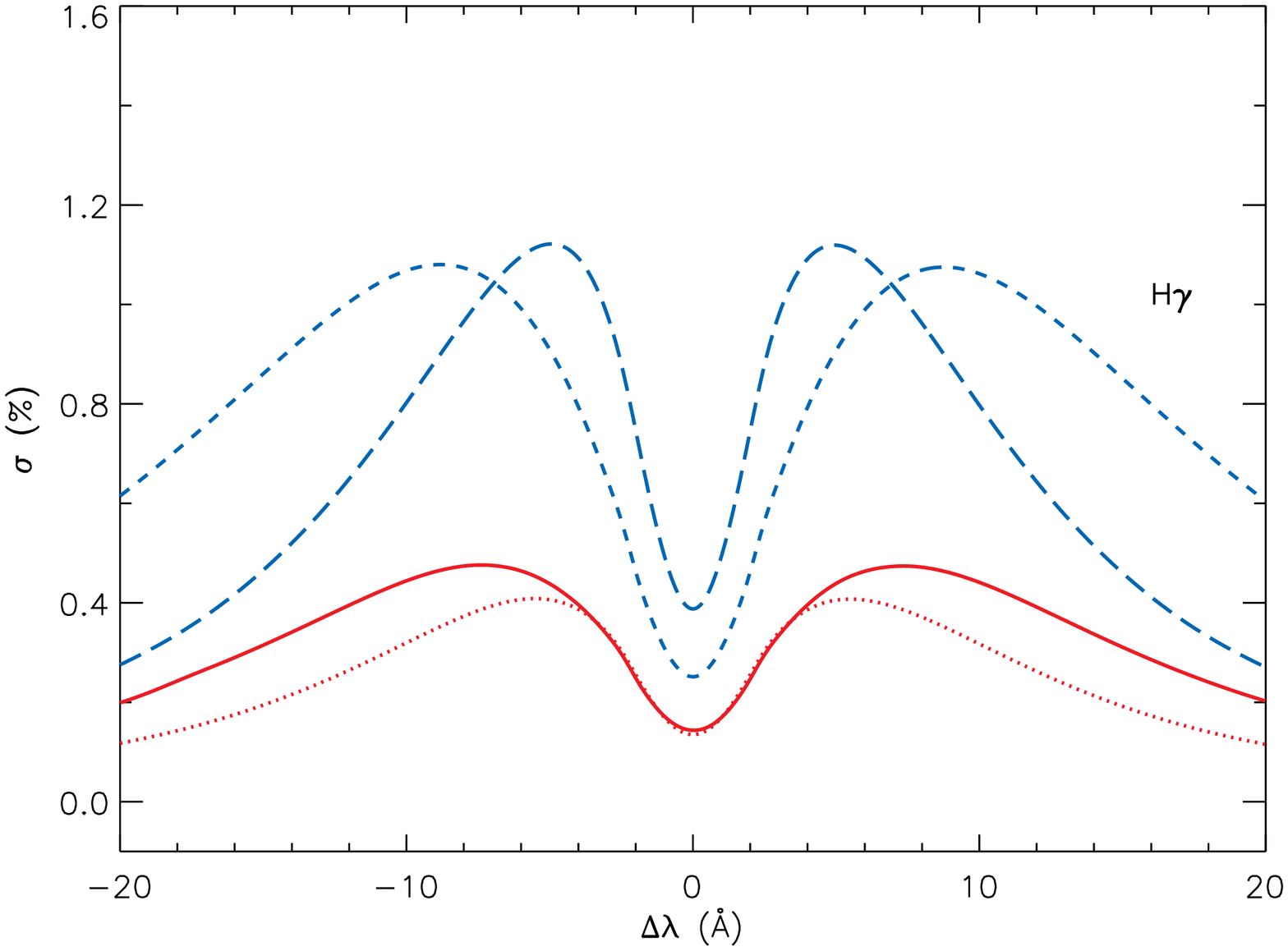}
\caption{The effective acceleration as a function of the Rosseland
optical depth for different rotation phases calculated for several magnetic field configurations and induced
\emf\ The resulting standard deviations around \hgamma\ line are shown in the
bottom panel.}
\label{fig:geff}
\end{figure*}

Not for all models considered in our study it was possible to fit the amplitude
of the observed standard deviation with the inward-directed Lorentz force.
This is true for models with $B_{\rm q}/B_{\rm d}>-1$, where the changes
in longitudinal magnetic field and magnetic field modulus result in very 
narrow phase-resolved variations in magnetic force term. As an example,
the left panel of Fig.~\ref{fig:geff} illustrates the run of the $g_{\rm eff}$
with the Rosseland optical depth in the atmosphere of \hd\ ($B_{\rm q}/B_{\rm d}=-2.5$)
for inward- and outward directed Lorentz forces computed under different
assumptions about induced electric field. An increase of the \emf\ value by a factor of two
considerably changes the amplitude of the effective gravity, but the computed
standard deviation does not change much (see lower left panel of Fig.~\ref{fig:geff}).
This is why even a large increase of \emf\ does not reveal itself in 
a standard deviation plot. Obviously, the situation may change once a more complex
geometry of the magnetic field is introduced, however it is connected with the
introduction of additional free parameters making the fitting procedure
ambiguous.
In contrast, the outward-directed magnetic force seems to have a larger impact
on the model pressure structure: varying \emf\ value from $1\times10^{-11}$~CGS to
$1.2\times10^{-11}$~CGS changes the amplitude of standard deviation by about
a factor of two or more.

The amplitude of the standard deviation around hydrogen lines strongly depends on the magnitude of the inward-directed Lorentz force for different magnetic field geometries.
For instance, the middle panel of Fig.~\ref{fig:geff} illustrates standard deviations
for the model with $B_{\rm q}/B_{\rm d}=-0.5$: changing induced \emf\ by a factor of two
considerably increases the amplitude of the standard deviation.
Similarly, the right panel of Fig.~\ref{fig:geff} shows the predicted variations 
for the purely dipolar model.

\subsection{Comparison with the observations}
\begin{figure*}
\includegraphics[width=0.33\hsize]{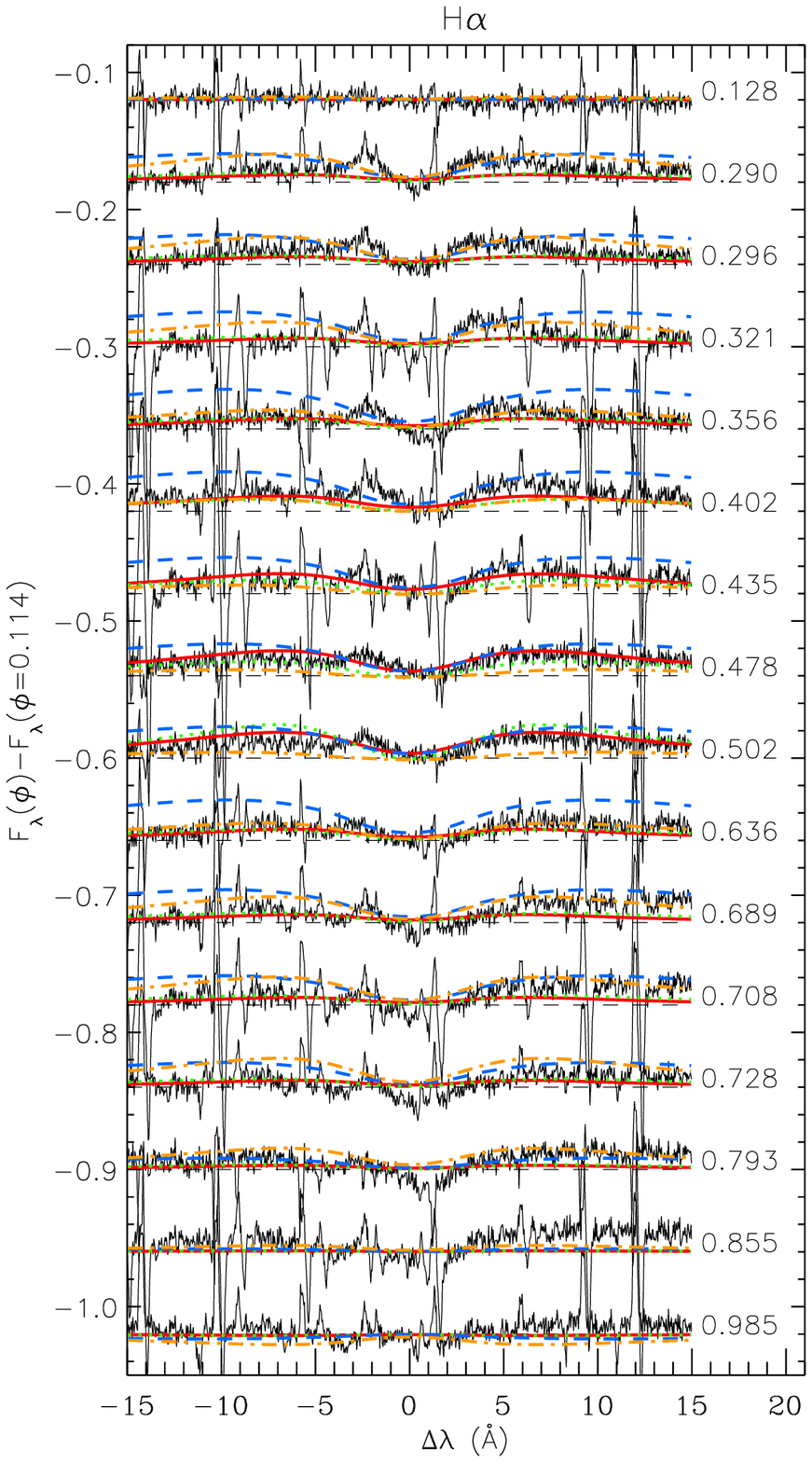}
\includegraphics[width=0.33\hsize]{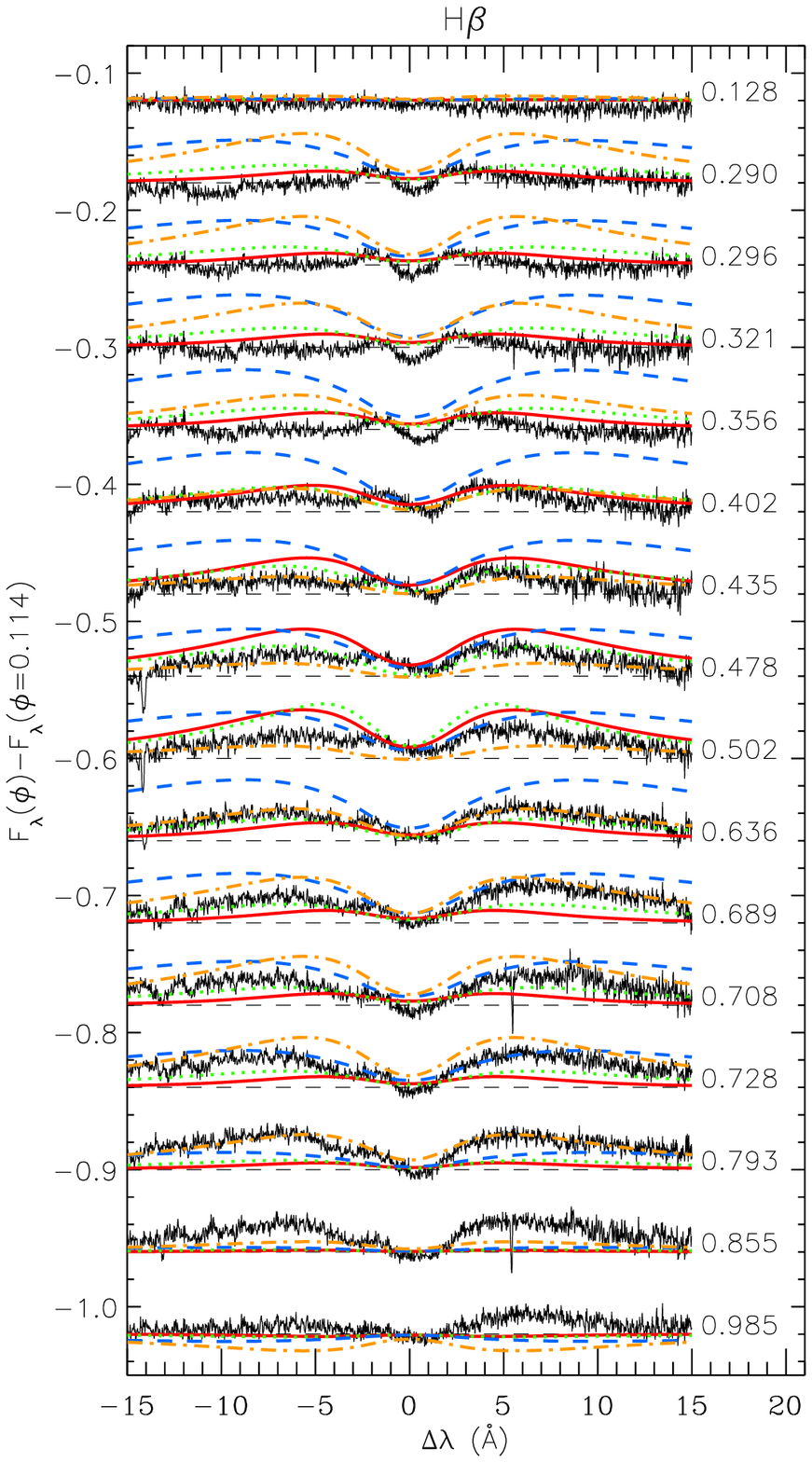}
\includegraphics[width=0.33\hsize]{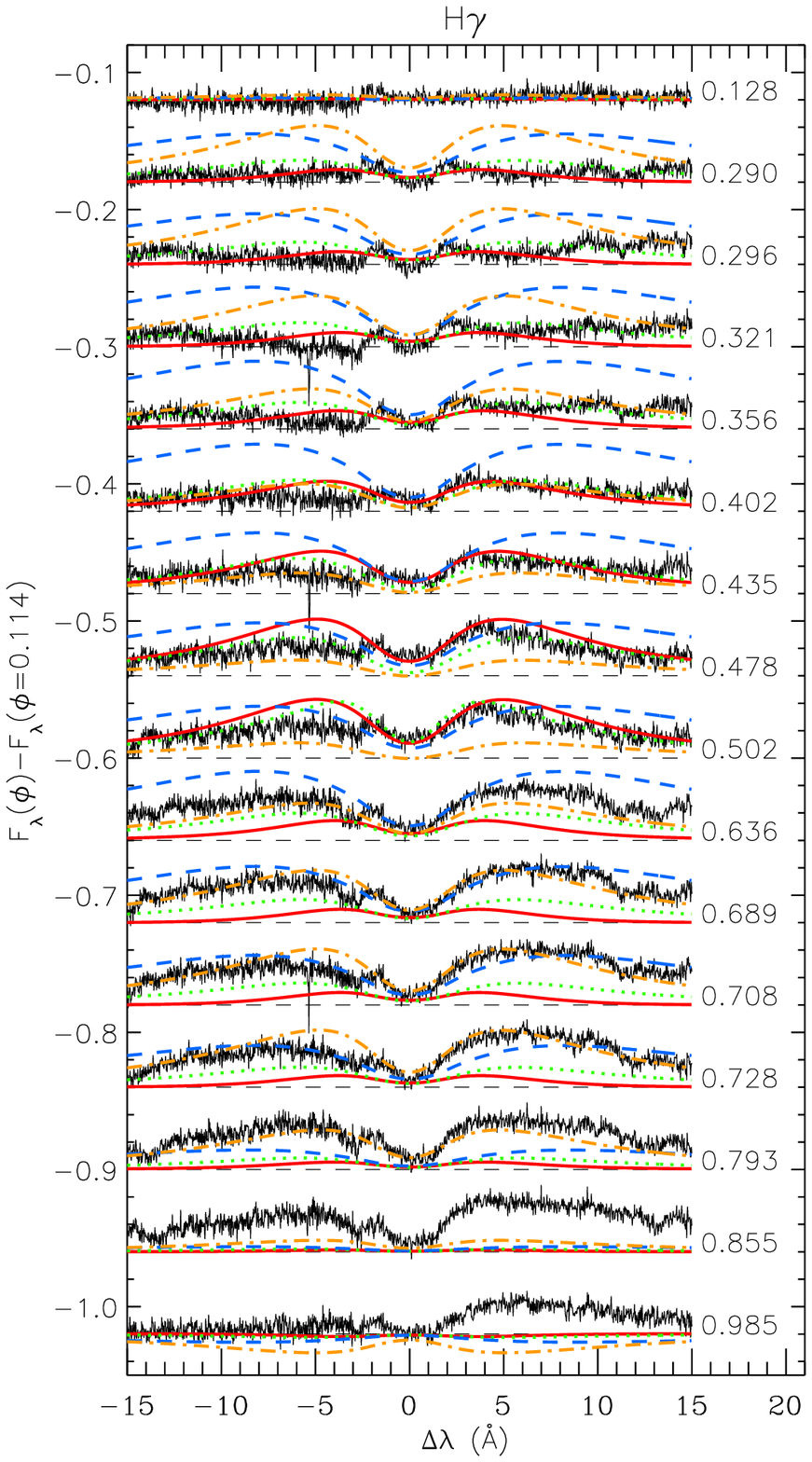}
\caption{Residual profiles of the \halpha, \hbeta, and \hgamma\ lines 
for each of the observed phase relative to the phase $0.114$. 
Thin solid line~--~observations, theoretical profiles are shown for the
following models: thick solid line~--~$B_{\rm q}/B_{\rm d}=-2.5$, $c_2/c_1=7$ model
(outward-directed Lorentz force), 
dotted line~--~$B_{\rm q}/B_{\rm d}=-4.5$, $c_2/c_1=7$ model (outward-directed Lorentz force),
dash-dotted line~--~$B_{\rm q}/B_{\rm d}=0.5$, $c_2/c_1=0.1$ model (outward-directed Lorentz force),
dashed line~--~$B_{\rm q}/B_{\rm d}=-1.0$, $c_2/c_1=2$ model (inward-directed Lorentz force).
The residual spectra for consecutive phases are shifted in the vertical
direction. The thin dashed line gives the zero level for each spectrum.}
\label{fig:resid}
\end{figure*}

In the following we compare the residual theoretical and observed
Balmer lines with model predictions based on different assumptions about
the magnetic field geometry and the direction of the Lorentz force. 
The residuals are obtained by subtracting a spectrum at a reference phase 
($\phi = 0.114$ where the Balmer profiles have the largest widths) 
from all the other spectra. Figure~\ref{fig:resid} illustrates residual 
H$\alpha$, H$\beta$, and H$\gamma$ line profiles
for each of the observed rotation phases. The positive sign of the residuals implies 
that the lines at the current phase are narrower
than those obtained at the reference phase. 
It is seen that the characteristic behaviour of hydrogen lines demonstrates
a single wave variation with the most noticeable effect at phases 
between $\phi=0.402$ and $\phi=0.855$. The effect is also seen in the red
wings of lines at $\phi=0.985$, however it is smeared out in the blue wing.
We emphasize that this systematic asymmetry, with the blue wing lying below the red one, 
is observed for all three studied Balmer lines. Inaccuracy of the spectrum processing 
can be one of the reasons for this effect.
However, our data reduction is identical to the analysis of $\theta$~Aur \citep[see ][]{lorentz-2007}, 
which shows no such asymmetry. Thus, we suspect that the asymmetry may also have physical origin due to
non-stationary, magnetically-channeled stellar wind from the surface of 56\,Ari. 
Very fast rotation and a relatively high temperature of this star make it plausible that its 
wind produces the variable, obscured P-Cyg feature distorting blue wings of the Balmer lines. 
The presence of a weak feature approximately $2.5$\AA\, blueward the line center seen for
all three hydrogen lines could also be an argument for the wind (for \hgamma\ line the presence of
this feature could alternatively be explained by absorption in \ion{Ti}{ii} and \ion{Fe}{ii} lines, however in case
of \hbeta\ there are no spectral lines that could contribute to this feature). 
Nevetheless, despite these problems introduced by an unknown physical process, the characteristic shape 
of the Lorentz force induced variability is clearly seen in the hydrogen lines of \hd, making it possible 
to perform the analysis in the framework of our modeling approach.

None of the tested theoretical models appeared to fit the observed variability
in all phases, but still generally describes the data more or less reasonably well for more than half of them
(see, for example, Model~1 or Model~4).
Varying the magnetic field geometry and the ratio of induced
dipolar and quadrupolar equatorial \emf's ($c_2/c_1$), we could achieve
a good agreement for a certain phase interval only: either it was possible to fit the
observations around phase $\phi=0.5$ or for other phases only.
As an example, in Fig.~\ref{fig:resid} we plot some of the theoretical
predictions for models that provide a more or less reasonable fit to the
observations. In particular, models with the outward-directed Lorentz force
are shown for the following configurations:
\begin{itemize}
\item[] $B_{\rm q}/B_{\rm d}=-2.5$, $c_1=1.2\times10^{-11}$~CGS, $c_2/c_1=7.0$ (Model~1),
\item[] $B_{\rm q}/B_{\rm d}=-4.5$, $c_1=1.2\times10^{-11}$~CGS, $c_2/c_1=7.0$ (Model~2),
\item[] $B_{\rm q}/B_{\rm d}=+0.5$, $c_1=9.0\times10^{-11}$~CGS, $c_2/c_1=0.1$ (Model~3).
\end{itemize}
Also, the model with the inward-directed Lorentz force is presented:
\begin{itemize}
\item[] $B_{\rm q}/B_{\rm d}=-1.0$, $c_1=5\times10^{-10}$~CGS, $c_2/c_1=2$ (Model~4)
\end{itemize}
For all plotted models the inclination angle $i=50^\circ$
was assumed. Taking $i=70^\circ$ does not change much the disc-integrated
parameters of the magnetic field and thus leads to essentially the same
picture of the hydrogen lines variation (see below).

Models 1 and 2 have the same parameters except the strength of quadrupolar
magnetic field component. They produce almost the same fit to the
observed variations, and in the same manner are not able to fit phases at
$\phi=0.636$ and above (Although Model~2 gives systematically a little bit 
better fit there). Note that models with $B_{\rm q}/B_{\rm d}<-1.5$ give
the same kind of the fit, but we do not plot them here to avoid overcrowding the plot.
At the same time, Model~3 seem to be a preferable one
for these phases, however it fails to fit observations at phases 
$\phi=0.435, 0.478, 0.502$, and gives enormously high effect at phases
$\phi=0.321$ and down to zero. This is true also for the Model~4 with
inward-directed Lorentz force. This model fits reasonably well
such phases as $\phi=0.689, 0.708$, but yields the line wings that are
generally too wide comparing with observed ones. Thus, of the two
possible directions of the Lorentz force in our 
model 
we consider an outward-directed Lorentz force as the more reasonable choice
to describe observations of \hd. Due to problems with telluric lines the
continuum normalization around \halpha\ line is substantially inaccurate
comparing to other lines
and it is not possible to distinguish between different models there.

\begin{figure*}
\centering
\includegraphics[width=0.23\hsize]{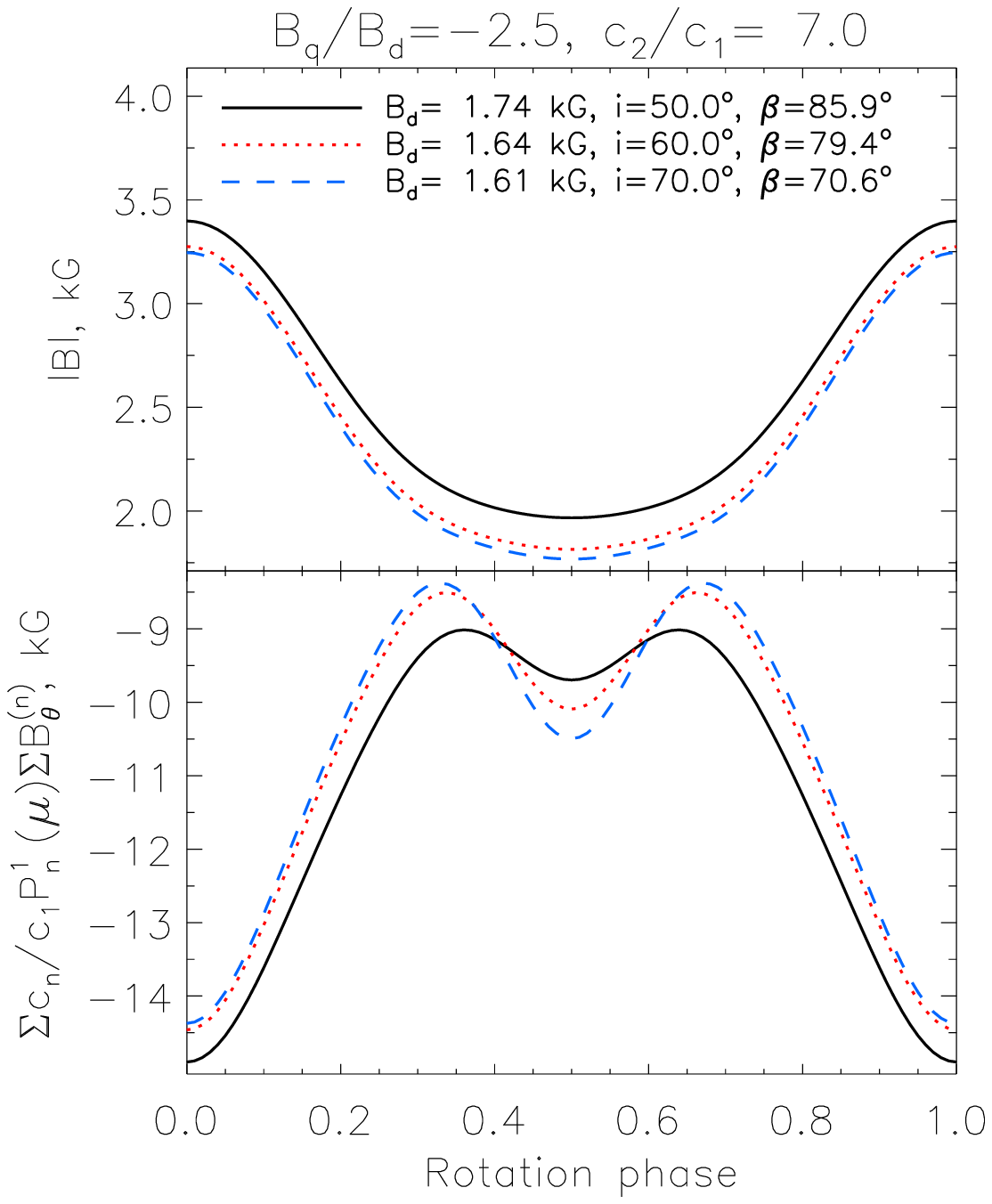} 
\includegraphics[width=0.23\hsize]{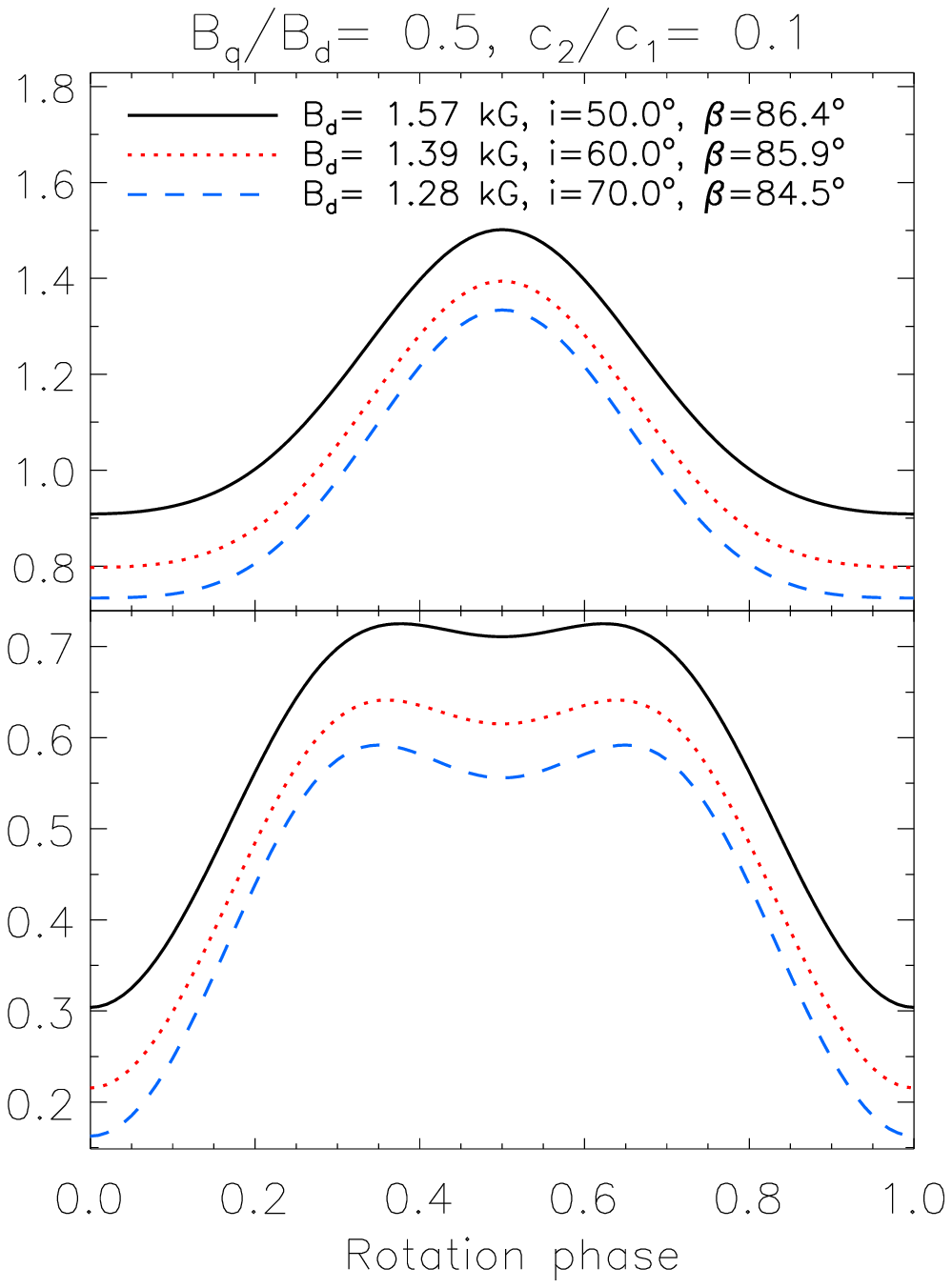} 
\includegraphics[width=0.23\hsize]{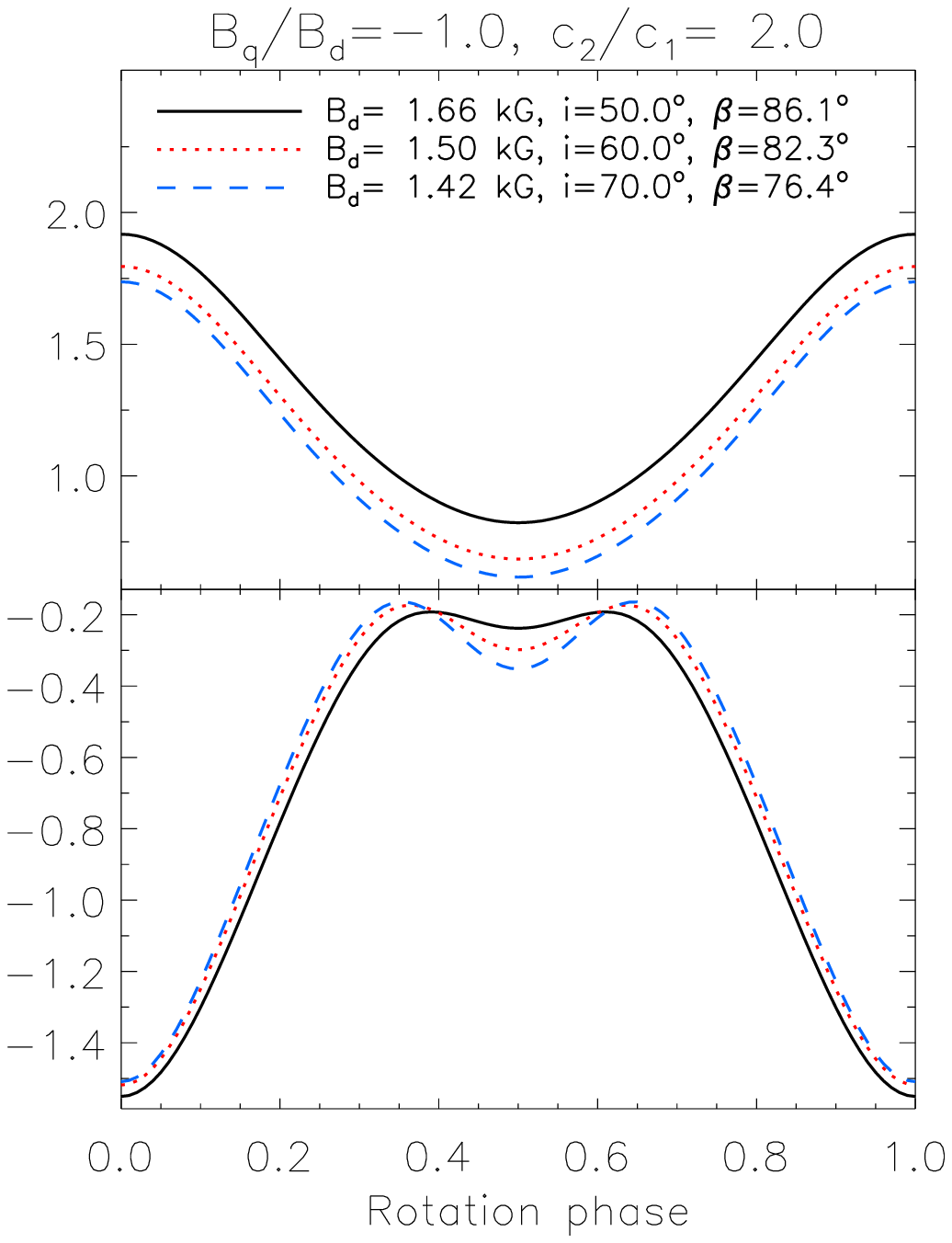} 
\includegraphics[width=0.23\hsize]{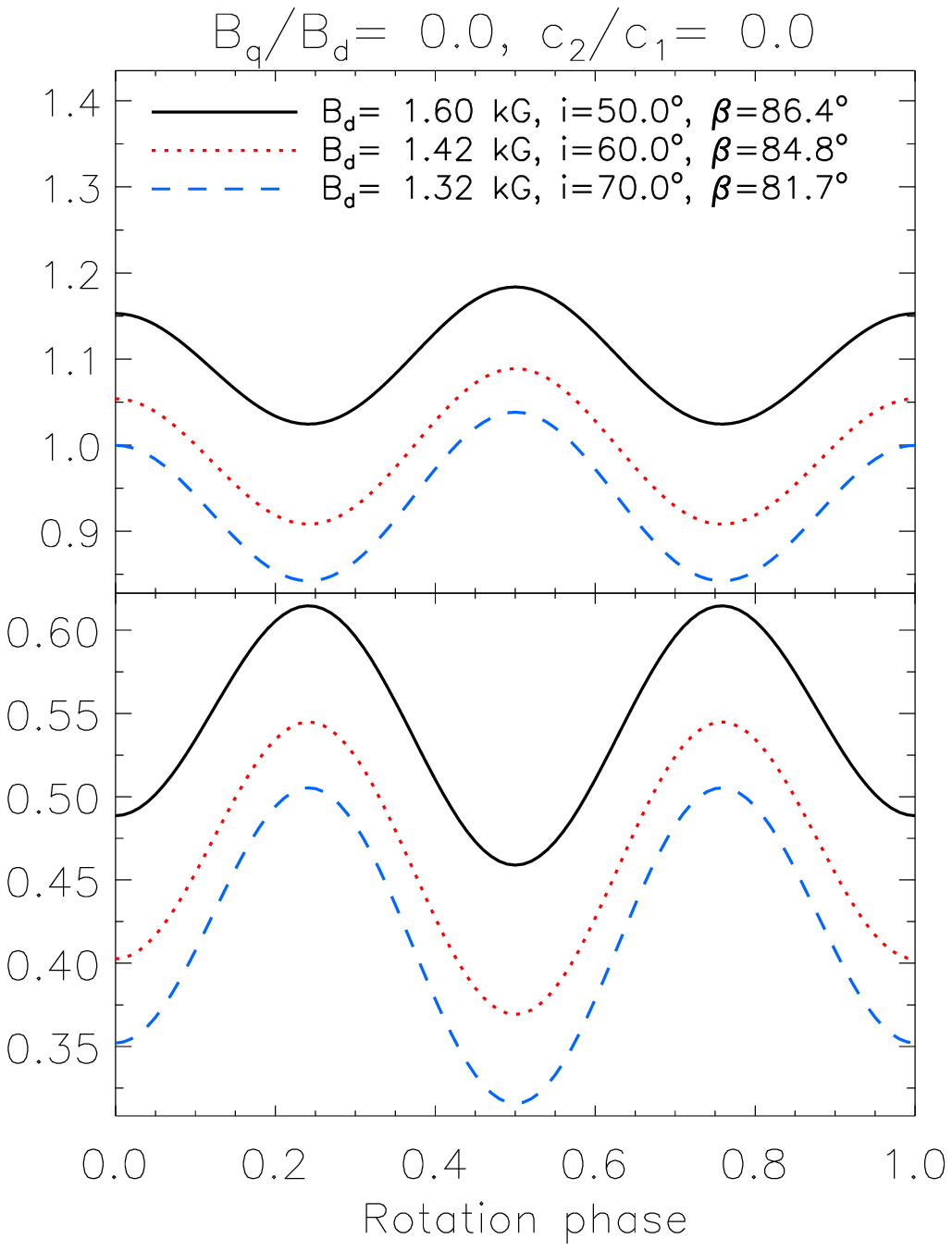}
\caption{Magnetic field modulus and Lorentz force parameter as a function 
of rotation phase for several magnetic field models.}
\label{fig:mag}
\end{figure*}

Testing models with different magnetic field parameters we tried to find
those that predict a single-wave variation of the magnetic force term over the
rotation cycle, as indicated by observations. Moreover, its run is likely to have a wide plato around
$\phi=0.5$ and drop rapidly close to $\phi=0$ and $\phi=1$ to fit observations
(see Fig.~\ref{fig:resid}). By varying the parameters $B_{\rm q}/B_{\rm d}$ and
$c_2/c_1$ (with the fixed $i$ and $\beta$) we succeeded to find 
sets of parameters that give this kind of plato, but in all cases it appears to be not 
as wide as it is needed to fit observations in all phases. This is illustrated in
Fig.~\ref{fig:mag} where we plot magnetic parameters used for the Lorentz force
calculation in some of the models mentioned above as a function of the rotation phase. 
The right panel in this figure illustrates predictions for the purely dipolar model. 
It is also seen that the inclination angle $i$ does
not play critical role in the present investigation: models with different
$i$ would give the same phase-resolved variation in hydrogen lines and any
amplitude difference between them can be adjusted by a proper choice of a $c_1$ parameter.

In this investigation we focused analysis on the hydrogen lines. It appears that they are most sensitive to the pressure changes introduced by the Lorentz force. As for metal lines, none of the strong \ion{Si}{ii} lines visible in the spectrum of \hd, $\lambda~5055.98~$\AA, $5466.48~$\AA,
$5466.89$~\AA, $6347.11$~\AA\, or $6371.37$~\AA, exhibits significant variation due to a non-zero Lorentz force. We have tested this by using the magnetic parameters 
of Model~1 (which has the largest amplitude of the $g_{\rm eff}$ variation, see Fig.~\ref{fig:geff}) and
recomputing spectrum models for every phase with the mean abundances.
We find no detectable changes in the line wings and less than $1$\% difference in the line cores between models with and without
Lorentz force. This difference is likely to be due to the differences in the
temperature distribution of these two models. No visible phase-dependent changes can be seen in the spectra corresponding to the models with Lorentz force. These results lead us to the conclusion that, due to their high pressure sensitivity
in the predominantly ionized plasma of the atmosphere of such relatively hot star, only the hydrogen lines are useful indicators for the 
magnetic pressure effects.

Similarly, we find no evidence for the influence of phase-dependent pressure effects on the stellar spectral energy distribution. The maximum difference between models with and without Lorentz
force is less than $2$\% in Balmer continuum. This corresponds to $\approx$\,0.01~mag difference in $c_{\rm 1}$ color-index and even less for
other Str\"omgren parameters. Thus, variation seen, for example, in the phase-resolved spectrophotometric scans of \hd\ published by \citet{adelman-phot} could not be attributed to the Lorentz force effects but are produced by inhomogeneous abundances and/or other mechanisms.

We note that a strong decrease of $g_{\rm eff}$ evident in Fig.~\ref{fig:geff} (up to $\approx$\,1~dex around $\log(\tauros)=0$)
compared to the non-magnetic case leads to a relatively small difference in the observed parameters due to 
a) the fact that this decrease does not affect the entire stellar atmosphere and 
b) non-local nature of the hydrostatic equation in the presence of depth-dependent $g_{\rm eff}$. 
The later implies that, for example, one order of magnitude 
increase of the magnetic gravity results only in three times lower gas pressure for the outward-directed Lorentz force, 
which is too small to significantly change 
the opacity coefficient and influence the model structure. The difference between magnetic models
for different rotational phases is even smaller since $g_{\rm eff}$ varies maximum by a factor of $\approx$\,2 for Model~1.

Finally, we stress that it is difficult to conclude anything with certainty regarding the preferable model of the magnetic
field geometry without additional accurate magnetic observations of the \hd.
Furthermore, other dynamic processes, such as Hall's currents and particle
diffusion, may contribute to the observed variations of hydrogen lines. 
These processes can not be accounted in our modeling due to their complex nature. 
Nevertheless, similar to the results of the \paperone, in this study we demonstrate that 
the observations can be described with a simple geometrical approach
under the assumption of strong surface electric currents in the atmosphere
of a main-sequence mCP star.

\section{Conclusions}
With the use of the high-resolution, phase-resolved observations of a magnetic CP star
\hd\ and employing modern model atmosphere technique we detected and investigated
variations of the Stark-broadened profiles of \halpha, \hbeta, and \hgamma\ lines
in a framework of a Lorentz force model.
Several proofs of the presence of significant magnetic pressure in the atmosphere
of \hd\ have been found:
\begin{itemize}
\item
The characteristic shape of the variation during a full rotation cycle of the star corresponds to those
described by \citet{kroll} and other authors 
as a result of the impact of a substantial Lorentz force (see \paperone\ and
references therein).
\item
Numerical calculations of the model atmospheres with individual abundances
demonstrate that the surface chemical spots 
cannot produce the observed variability in the hydrogen line profiles of the star.
\item
Our model shows a reasonable agreement with the observations
if the outward-directed magnetic force is applied assuming the
dipole+quadrupole magnetic field configuration. Unfortunately, due to large
uncertainties in the available observations of the longitudinal magnetic field 
it was not possible to conclude confidently about the strengths of the 
quadrupolar component.
\item
Taking into account a variety of possible solutions, we find that, to fit the amplitude
of a phase-resolved variation in \halpha, \hbeta, and \hgamma\ lines, the magnitude
of an induced equatorial \emf\ must be in the range $10^{-11}-10^{-10}$~CGS in case of
outward-directed Lorentz force and $\sim5\times10^{-10}$ in case of inward-directed
one.
\end{itemize}

\section{Discussion}
\hd\ is the second magnetic CP star for which we detected the characteristic variation
of the hydrogen Balmer line profiles and performed detailed modeling of the
Lorentz force effect. Our previous target, A0p star $\theta$~Aur, also demonstrated
significant variability in the hydrogen Balmer lines (see \paperone).
However, in the case of $\theta$~Aur, \citet{borra} provided a much more accurate measurements
of the longitudinal magnetic field variations, which allowed us to
determine more precisely geometry of its surface magnetic field. Nevertheless, for both stars we find that 
the outward-directed Lorentz force is needed to explain
observations. However the induced \emf\ for \hd\ may
differ by an order of magnitude, but this remains uncertain due to poorly known magnetic
field geometry.

A single-wave variation of the residual spectra is a characteristic signature
for both $\theta$~Aur and \hd. Since both stars have high inclination angles and magnetic obliquities, in the framework of our Lorentz force model this variation 
indicates the presence of a more complex magnetic field geometry
than a simple dipole. Such a variation can be obtained in the non-dipolar theoretical models by a
proper choice of induced \emf\ for each of the multipolar components (e.g. the $c_2/c_1$ ratio in the case
of dipole$+$quadrupole combination).
Furthermore, for both stars the amplitude of the longitudinal magnetic field
variation is about $\approx500$~G,
which can be the reason for similar
amplitude of the detected Balmer lines variation ($\approx1$\%) since the effective 
temperatures of stars are different ($\teff$($\theta$~Aur)$=10\,400$~K, $\teff$(\hd)$=12\,800$~K). 

Similar to the \paperone, we do not consider here any details about the physical
mechanisms that could be responsible for the observed Lorentz force. The final conclusion
about the nature of the significant magnetic pressure can only be obtained when more
sophisticated models of the magnetic field evolution and its interaction with highly
magnetized atmospheric structure will become available and/or alternative models
will be tested \citep[however, for some of the estimates see discussion in ][]{lorentz-2007}.

In the present work we made use of a simple
geometrical 1D model of Lorentz force: the surface averaged values of the transverse
magnetic field and the magnetic field modulus are introduced in the hydrostatic
equation of the stellar matter. Future investigations can benefit from taking into account 2D effects with direct
surface integration of the hydrogen line profiles computed with individual
models. This could probably also open a possibility
to account for the Hall's currents. 
Unfortunately, as it was mentioned above, this is difficult to do at present, 
but of no way impossible once more computational resources become available.

The dependence of the observed variability in hydrogen lines upon the magnetic field
geometry and strength is one of the key element in our investigation. If such a dependence
exists, it could bring a number of theoretical constrains about the interaction of
the magnetic field with stellar plasma. So far, we have analyzed only
two stars with occasionally similar longitudinal magnetic field intensity.
Note that we are limited to stars for which the configuration of a magnetic field
can be extracted from the literature and which
can be observed with highly stable spectrometers like BOES in order to reduce possible
errors in spectra processing. Thus, observations of other mCP stars 
are needed to conclude about the connection between magnetic 
field and variability seen in Balmer lines.

\begin{acknowledgements}
The authors are greatly thankful to Tanya Ryabchikova for her help with the preparation
of line lists used in DI. We also acknowledge the use of a cluster facilities at
Institute of Astronomy, Vienna University. This work was supported by FWF Lise Meitner grant Nr. M998-N16 to DS. 
OK is a Royal Swedish Academy of Sciences Research Fellow supported by a grant from the Knut and Alice Wallenberg Foundation.
Han acknowledges the support for this work from the Korea Foundation for 
International Cooperation of Science and Technology (KICOS) through grant No. 07-179.
Based on INES data from the IUE satellite.
\end{acknowledgements}

\end{document}

%% file: common-def.tex
\def\llm{{\sc LLmodels}}

\def\logg{\log(g)}
\def\teff{T_{\rm eff}}
\def\tauros{\tau_{\rm Ross}}

\def\kms{km\,s$^{-1}$}
\def\bs{$\langle \vec{B} \rangle$}
\def\paperone{Paper~I}
\def\halpha{H$\alpha$}
\def\hbeta{H$\beta$}
\def\hgamma{H$\gamma$}

\def\synth{{\sc SYNTH3}}

\def\bz{$\langle B_{\rm z} \rangle$}
\def\emf{$e.m.f.$}
\def\vsini{$v\sin i$}

%% file: tables/tab-obs.tex
\begin{table}
\caption{Observations of \hd.}
\centering
\begin{tabular}{ccc}
\hline\hline
No. & JD & Rotation Phase\\
\hline
 1 & 2453250.2191 &  0.321 \\
 2 & 2453250.3017 &  0.435 \\
 3 & 2453251.2284 &  0.708 \\
 4 & 2453251.3359 &  0.855 \\
 5 & 2453306.1282 &  0.128 \\
 6 & 2453306.2942 &  0.356 \\
 7 & 2453308.0681 &  0.793 \\
 8 & 2453309.0295 &  0.114 \\
 9 & 2453309.1620 &  0.296 \\
10 & 2453666.0880 &  0.636 \\
11 & 2453671.0690 &  0.478 \\
12 & 2453671.0860 &  0.502 \\
13 & 2453759.0098 &  0.290 \\
14 & 2453760.0284 &  0.689 \\
15 & 2453760.9714 &  0.985 \\
16 & 2453762.0026 &  0.402 \\
17 & 2453762.9682 &  0.728 \\
\hline
\end{tabular}
\label{tab:obs}
\end{table}

%% file: tables/tab-abn.tex
\begin{table}
\caption{Abundances (in $\log N_{\rm el}/N_{\rm total}$)
of \hd, used for determination of model atmosphere
parameters.}
\centering
\begin{tabular}{cccccc}
\hline\hline
       & He    & Mg    & Al    & Si    & Fe    \\
\hline
\hd\,  & -2.10 & -5.51 & -6.17 & -3.53 & -4.09 \\
Sun    & -1.10 & -4.51 & -5.67 & -4.53 & -4.59 \\
\hline
\end{tabular}
\label{tab:abn}
\end{table}